\documentclass[letterpaper, 10 pt, conference]{ieeeconf}  

\IEEEoverridecommandlockouts                              

\usepackage{balance} 

\usepackage{array}
\usepackage{booktabs}
\usepackage{graphicx}

\usepackage[table]{xcolor}
\usepackage{colortbl}
\usepackage{booktabs}

\definecolor{gray1}{gray}{0.95} 
\definecolor{gray2}{gray}{0.90} 
\usepackage[hidelinks]{hyperref}

\title{\LARGE \bf
Can You See How I Learn? Human Observers' Inferences about Robot Learning Processes
}

\author{Bernhard Hilpert$^{1, 2}$, Muhan Hou$^{2}$, Kim Baraka$^{2}$ and Joost Broekens$^{1}$
\thanks{$^{1}$LIACS,
         Leiden University, Leiden, The Netherlands
         }%
\thanks{$^{2}$
         VU Amsterdam, Amsterdam, The Netherlands
         }%
}

\begin{document}

\maketitle
\thispagestyle{empty}
\pagestyle{empty}

\begin{abstract}

Interactive robot learners often exhibit learning behaviors that are not intuitively interpretable by human observers, which can result in suboptimal feedback in collaborative teaching settings. 
Yet, how humans perceive and interpret robot learning behavior is largely unknown. In a bottom-up approach with two experiments, this work provides a data-driven overview of the factors influencing human observers' understanding of the robot learning process. 
A novel, observation-based paradigm to directly assess human inferences about robot learning was developed.
In an exploratory interview study (\textit{N}=9), four common themes in human inferences were identified: Agent Goals, Knowledge, Decision Making, and Learning Mechanisms.
A second confirmatory study (\textit{N}=34) applied an expanded version of the paradigm across two tasks (navigation/manipulation) and two Reinforcement Learning (RL) algorithms (tabular/function approximation). Analyses of 816 responses confirmed the reliability of the paradigm and refined the thematic framework, revealing how these themes evolve over time and interrelate. 
Our findings provide a human-centered understanding of how people make sense of robot learning, offering actionable insights for designing interpretable RL systems and improving transparency in Human-Robot Interaction (HRI).

\end{abstract}

\section{Introduction}
Hybrid Intelligence (HI) refers to collaborative human–machine systems that combine strengths to accomplish tasks neither could perform alone \cite{akata2020research}.
A key example is Human-Interactive Robot Learning (HIRL), where human teachers guide learning robots using signals such as feedback or demonstrations \cite{baraka_humaninteractiverobotlearning_2026}.
While humans can provide insight, expert knowledge, or common sense \cite{celemin2022interactive}, feedback is often suboptimal: delayed, inconsistent or based on misinterpretation of the robot's behavior \cite{arzate2020survey}.
Recent surveys emphasize explainability and human involvement across all stages of the HIRL lifecycle 
\cite{retzlaff2024human}. However, 
while mental models \cite{bansal2019beyond} and their mismatches \cite{richter2024reducing} have been a key area of research in HRI, existing literature predominantly examines teacher inferences about robot learning indirectly via interaction patterns \cite{ho2019people} or simulated settings \cite{grislain2023utility}.
Thus, existing paradigms confound observation with interaction: when humans actively teach, their expectations shape not only the feedback but also the robot's behavior, making it hard to isolate what they actually understand \cite{ho2019people}.
This may obstruct an unbiased examination on the human's perspective of the learning progress, especially at later learning stages. 
Open questions include the correct setting, how to even query inferences about a robot's learning process from human observers, and which aspects of learning humans attend to when making sense of robot behavior.

The main Research Question for this study therefore is: \textit{What do human teachers infer about RL-based robot learning processes from observing the robot's behavior?}
To investigate how observers naturally conceptualize robot learning, this work takes a human-centered, bottom-up approach, resulting in two main contributions: 
1) a novel observation-only experimental paradigm designed to isolate and directly elicit intuitive observer inferences about robot behavior, 2) a granular, empirical framework identifying four recurring inference themes (Goals, Knowledge, Decision Making, and Learning Mechanisms) with 16 subclusters, empirically collected across two experiments, spanning different tasks and Reinforcement Learning (RL) algorithms. 
A detailed analysis further examines how they evolve over time and contribute to observers’ understanding of robot learning\footnote{A preliminary version was presented as an extended abstract at AAMAS 2025 and as a non-archival paper at the AAMAS ALA Workshop 2025 (available as a preprint on arXiv).
}.

\section{Background and Related Work}
The mechanisms of RL are inherently difficult for human observers to interpret from behavior alone, due to algorithmic complexity \cite{habibian2023review} and the challenge of intuitively interpreting concepts like reward functions \cite{hejna2023few}.
In interactive settings, a teacher's perception of a RL-based robot's behavior can recursively shape their interpretation of the learning process and thereby influence the course of the interaction \cite{macglashan2017interactive}.
For instance, \cite{ho2019people} demonstrate that teachers are strongly inclined to use rewards in the teaching process intuitively as communicative signals rather than a reinforcement signal,
indicating a persistent misalignment between users’ assumptions and actual RL mechanisms.
Although RL agents are designed to optimize cumulative rewards, the observed interaction pattern suggests that human teachers seem to interpret their actions as if the agent were responding to the communicative intent rather than reinforcement signals.
The authors explained this through assumed differences in how teachers interpret the agent’s state space, and that teachers seem to evaluate the agent by comparing its inferred policy to their own desired policy \cite{ho2019people}.
While such judgments may reflect a reasonable teaching strategy from the human perspective, the formation of positive reward cycles in the teaching process based on the resulting feedback \cite{ho2019people} highlights a disconnect between human interpretations of learning and the underlying mechanisms of RL.

Further work indicates that these communicative assumptions about feedback generalize to Human-Human Interaction, where punishment and reward are also interpreted through inference \cite{sarin2021punishment}.
More broadly, the Inferential Social Learning framework describes behavior interpretation as a process of "probabilistic inferences, guided by an intuitive understanding about how people plan and act" \cite{gweon2021inferential}. This suggests that individuals acquire knowledge about their interaction partners by inferring insights from behavior observation. 
Understanding how this process applies in the context of Human-Agent Interaction (HAI) settings,
is therefore crucial in order to design effective HIRL settings.
However, research in HAI has primarily studied teacher inferences indirectly, by analyzing interaction behaviors. The inferences themselves have rarely been studied directly.
The present work addresses this gap by directly examining how human observers infer and interpret the learning processes of RL-based robots from behavior observation alone.

\section{Exploratory Interview Study}
In this first exploratory study, an observational experimental paradigm is developed and applied, through which this question is studied 
and its exploratory results presented.

\subsection{Methodology}
\subsubsection{Participants} 
A total of \textit{n}=9 (\textit{n}=3 female, \textit{n}=6 male) were recruited. Five were bachelor students with RL exposure, four had no RL background.

\subsubsection{Set-up \& Procedure}
In the absence of a validated paradigm to directly examine observer inferences about RL processes (as a proxy for HIRL), an adapted version of the Grounded Theory framework was used to develop a reliable experimental procedure \cite{glaser2017discovery}. 
An exploratory qualitative interview setting was selected, due to the suitability of this method for studying undirected perceptual data \cite{silverman2017you}. 
Participants observed an RL agent’s learning process with a framing of being a teacher, but no possibility for intervention.
Firstly, they received a short introduction to RL together with an example video similar to the later stimuli 
and were asked to imagine helping the agent learn the task by teaching. 
Secondly, they were presented with videos of an RL agent learning to solve a task and were asked to share their observations about the learning process in a think aloud protocol \cite{oh2009think} to elicit a verbal report that is as undirected, undisturbed and constant as possible \cite{boren2000thinking}. 
Interviews (ca. 45 minutes) were conducted in person, audio-recorded; recordings were erased after an anonymized transcription. 
A written consent for this procedure was obtained from the participants beforehand;  
they were debriefed after the experiment. 

\subsubsection{Stimuli}A tabular Q-learning agent with Temporal Difference update was implemented.  
Stimuli consisted of rendered recordings of this agent solving the default 4x4 and 8x8 grids of the OpenAI gym \textit{FrozenLake} environment \cite{brockman2016openai}. 
To support coherent perception of the learning process, but also assess whether chunking affects richness of observations, two presentation formats were used. One group viewed two full-length videos (entire learning process); the other viewed six shorter clips, each showing a third of the process. Participants were balanced across these conditions and RL backgrounds (2 per condition).

\subsubsection{Measurements} 
Participants were instructed to explicitly report on the agent’s learning process (rather than 
behavior) while watching the videos. 
With no validated questionnaire available, a pilot session with an RL expert 
guided question phrasing in line with Grounded Theory principles 
\cite{glaser2017discovery}, specifically formulations that queried most sensibly the human inferences about the learning process.
Based on their input, two open-ended questions were conceived: "\textit{What do you think is going on in the brain of the agent?}" and "\textit{Why do you think the agent is doing what it is doing?}"

\subsubsection{Analysis} 
Audio recordings were transcribed and participant answers were pooled per video, to ensure full anonymization.
Individual statements were then isolated to enable systematic analysis of the think-aloud data.
Then, an 
inductive, reflexive thematic analysis was conducted based on an adaptation of \cite{braun_thematicanalysispractical_2021}, following a three-phase pipeline: 
(1) Data Reduction: condensing statements into core semantic units and grouping by synonymy; (2) Theme Generation: clustering core meanings into latent themes and assigning descriptive labels; and (3) Refinement: cross validating themes against the full dataset to ensure internal consistency.
To maintain conceptual continuity across datasets, primary coding was executed by a single investigator, following the recommendation of \cite{braun_thematicanalysispractical_2021}. To establish systemic trustworthiness and counteract individual coder bias, the evolving thematic structure was subjected to 30 hours of systematic peer-debriefing and critical review with senior HRI researchers, followed by an external validation workshop ($N=26$) to verify the accuracy of the generated framework.

\subsection{Results}
Data collection with the presented paradigm resulted in 264 isolated statements. 
Thematic analysis revealed four common emergent themes in observers' inferences about agent learning processes. 

\textbf{Agent Goals: }includes 23 inferences about the learning agent's goals.
Examples include "\textit{The agent wants to explore and understand the environment that it is in}", "\textit{It is trying to avoid the ice}", "\textit{It is trying to find a way that it can navigate the environment as quick as possible}" and "\textit{seems not to realize what the idea is}".

\textbf{Agent Knowledge: }
inferences that the agent’s learning process involved acquiring and using knowledge about the environment or task (36 mentions).
Examples include "\textit{It seems to have gathered some sort of sense of the second row being a bad row}", "\textit{it remembers, “alright I was in that tile and then I moved to that tile and there was a lake right there"}", "\textit{he knows exactly which tiles to not step onto, because he memorized it}".

\textbf{Agent Decision Making: }
inferences about how the agent takes decisions during learning, including individual actions and overall policies (50 mentions).
Examples include "\textit{It is just trying out different moves}
, "\textit{He follows a system: 1 down, one right, then 2 down one right, then three down}" and "\textit{He realized he cannot take the first second rows and thinks whether he can take the third}".

\textbf{Agent Learning Mechanisms: }
inferences about mechanisms by which the agent updated its internal representations 
(18 mentions).
Examples include "\textit{it is just mapping certain actions to certain states}", "\textit{He has moved two tiles to the right in the first try and realized it was not a mistake}".

\textbf{Other: }includes additional themes that were too infrequent to be classified as separate themes like perception (e.g., vision), feelings, communicative signals. 

\subsection{Discussion}
This experiment introduced a new observational paradigm to examine observer inferences of agent’s learning process from its behavior.
A data-driven analysis revealed four recurring themes in these inferences — Agent Goals, Knowledge, Decision Making, and Learning Mechanisms — offering insight into how observers made sense of learning agents over time. 

Agent Goals emerging as a distinctive theme indicates that task understanding plays a key role in observer inferences of the learning process. 
It reveals observer assumptions about the agent's understanding of the task structure, but could also reflect the observers' own interpretation of the task and environment. 
This is especially important, as every user approaches each task with their own set of assumptions, 
which can shape or even distort their interpretation of learning.
For instance, inferring that the agent aims to reach the exit “as safely as possible,” despite safety not being encoded in the reward structure. 
This highlights a broader challenge in human-agent interaction, where users project task goals that may misalign with an agent’s actual goals.

The emergence of Agent Knowledge indicates that observers inferred some form of internal representation of its past experiences or task structure. 
Unlike Goals, knowledge is a latent, less directly observable concept, yet observers appeared to attribute it readily. 
This indicates a more cognitively rich, anthropomorphized mental model of the agent, using assumed knowledge to explain how learning unfolds.

The emergence of Agent Decision Making indicates that observers constructed a procedural view of the agent’s learning process, beyond only structural components like goals and knowledge. 
Rather than seeing the agent as passively reacting to stimuli, observers infer it actively weighing options, selecting actions, or following strategies.
This suggests that observers interpret learning not just as an internal update, but as an active process of deliberation with a sense of control
, similar to human-like decision making.
This may hint at a potential over-attribution of deliberate strategy to the agent’s processes.

The emergence of Learning Mechanisms suggests that observers inferred not just behavior, but processes of adaptation, seeing them as actively learning entities.

The analysis did not reveal any differences in inferences that were based on demographic variables. 

Together, the four themes suggest that observers employ an integrated, procedural view of agent learning, that appears to be anthropomorphized, aligning with prior findings in cognitive science \cite{gweon2021inferential, ho2019people} and Human-Agent studies \cite{immel2025patients, hilpert2021employing}, where the attribution of known concepts made complex behavior more interpretable.

\textbf{Limitations and Lessons Learned}

Given the small sample size and exploratory nature of the study,  
the findings should be viewed as preliminary. 
A larger-scale study is needed to validate these initial results in a more systematic and robust manner.  
Secondly, participants in the split-up video condition produced approximately 3 times as many statements as those in the full-length video condition, suggesting that this format supports more differentiated and richer inferences. 
To expand the broad concepts found so far, the next step requires refining the experimental paradigm with a more structured, differentiated questioning format.
Next, the framework proved to be highly anthropomorphized. This could in part be due to the human tendency to apply pre-existing knowledge to virtual agents to facilitate understanding \cite{schneeberger2019can}, but also due to the stimulus material (i.e. the human-like agent in FrozenLake) and the question formulation (i.e. referring to the agent's brain processes). A next step should employ
a modification of the paradigm with less anthropomorphized, robot-centric stimuli and questioning. 
Lastly, this study examined observer inferences about an agent's learning process only for a very simple tabular agent. Since most interactive teaching settings in robotic applications employ more sophisticated forms of RL, future work should also validate this framework for other types of agents.

\section{Confirmatory Questionnaire Study}
Building on the exploratory findings of Experiment 1, this second experiment aimed to confirm those results while explicitly extending their generalizability in a larger-scale study. Specifically, it pursued three primary objectives: 
\begin{enumerate}
    \item Validate the initial findings on a larger sample 
    \item Add granularity to the four inference themes 
    \item Generalize findings by expanding the experimental paradigm to 
    other RL algorithms and tasks
\end{enumerate}

To address these objectives, a larger-scale confirmatory questionnaire study was designed and implemented\footnote{ Data collection was conducted jointly with \cite{hilpert2026toss}}.

\subsection{Methodology}
In line with open research practices \cite{wessler2021empirical}, this experiment was preregistered and all materials (Questionnaire, instructions, stimuli, annotated data corpus and the preregistration protocols) are openly available on the OSF\footnote{\url{https://osf.io/fumd8/?view_only=9cec60dccbd446f08bd818d0b3612705}}.

\subsubsection{Participants} 
\vspace{1em}
A total of 35 participants was recruited through Prolific of which one was excluded due to inattentive responses, resulting in a final sample of
\textit{N} = 34 (26 Male, 8 Female, 
0 Non-binary, self-assigned or undisclosed).
Inclusion criteria required participants to be 18 years or older, with a minimum of 
95\% approval rate on Prolific, 100 completed tasks, and self-reported fluency in English, confirmed by country of residence.
\textit{N} = 34 (26 Male, 8 Female, 
0 Non-binary, self-assigned or undisclosed).
The participants ranged from 19 to 65 years old (\textit{M}\,=\,38.79 years, \textit{SD}\,=\,10.68 years). 23 participants owned a pet, 20 participants reported to have children. Participants received on average £10.42. 

\begin{figure}[h]
\centering
  \includegraphics[width=\columnwidth]{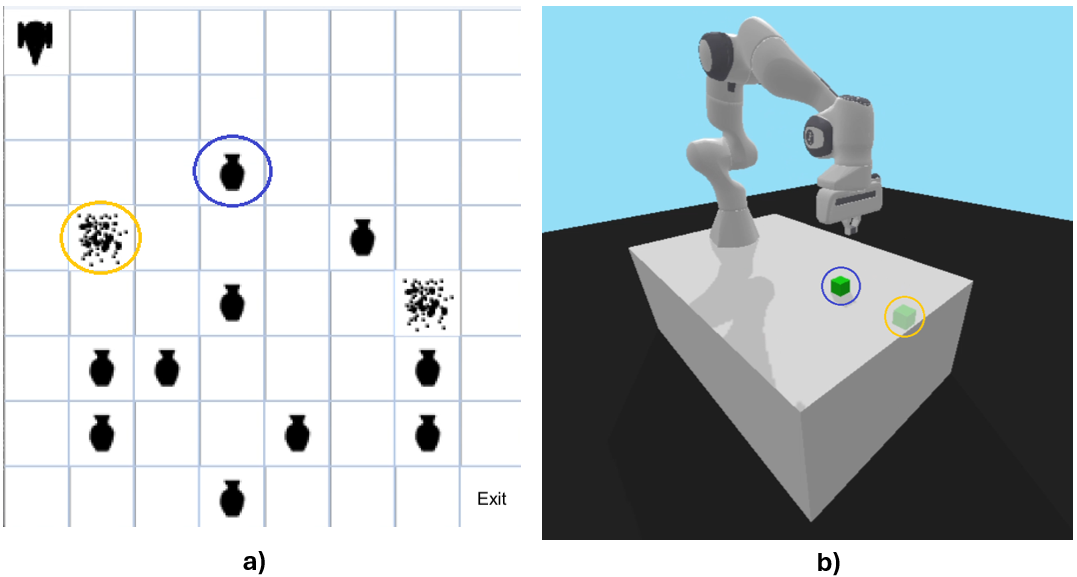}
  \caption{a): NT: robot needs to clean dirt (yellow) and avoid breaking furniture (blue) b) MT: the assistive robot needs to push "medication" (blue) to target position (yellow)}
  \label{fig:Tasks}
\end{figure}

\subsubsection{Expanded experimental paradigm}
\label{Paradigm}
To test the generalizability of the findings, the experimental paradigm was expanded in two key ways. First, a second task involving a Deep Deterministic Policy Gradient (DDPG; an off-policy, actor-critic algorithm for continuous action spaces) agent was introduced, to assess inferences about agent learning extend to function approximation settings. Second, to reduce anthropomorphism and increase ecological validity, both tasks were reframed as service robots performing everyday activities (see Figure \ref{fig:Tasks}).

In the navigation task (NT), the original tabular Q-learning agent operated in a modified 8x8 FrozenLake environment. Visual elements were adapted to depict a cleaning robot navigating a room: the elf was replaced with a conic robot icon (indicating orientation), ice tiles with white floor tiles, holes became vases, and the goal tile marked with an "exit" label. Two dynamic dirt patches were added, granting the agent an additional reward (+1) upon visit and disappearing afterwards, providing visible milestones to support observer inferences. Task logic and algorithm remained unchanged.

The manipulation task (MT) featured a DDPG agent framed as an assistive robot pushing a box of medication across a table to a target zone (where a patient that has trouble reaching could more easily pick it up). This was implemented using a modified Push task from the PandaGym environment \cite{gallouedec2021pandagym}, with fixed object/target positions near the table's edge. The reward function combined distance-based shaping rewards with a large terminal reward (+1000) upon successful delivery.

\subsubsection{Set-up \& Procedure}
Participants observed the RL agents' learning in a non-interactive, teaching framing. After recruitment via Prolific, they were directed to a study landing page outlining the procedure, ethical rights, and withdrawal options, followed by an informed consent form and commitment check.
Next, participants then received a lay-level introduction to RL concepts, designed to ensure baseline comprehension.\footnote{Instructions, questionnaire and stimuli were piloted with \textit{N=}4 test participants before recruiting in order to ensure understandability} They were then presented with two task scenarios (see Section~\ref{Paradigm}), each comprising three videos of a "robot’s" learning trajectory (see Section~\ref{Stimuli}), for a total of six videos.
Following each video, participants completed a modular block of questions (see Section~\ref{Measurements}, median: 57 minutes). 
After this, participants were redirected to Prolific and received compensation within 24 hours.
The study was approved by the university's ethics board.

\subsubsection{Stimuli}
    \label{Stimuli}

To reflect genuine learning processes, stimuli were generated by training each robot to convergence (defined as 5 consecutive successful episodes) and rendering roll-outs from saved checkpoint policies. 
For the NT, the agent was trained for 10000 episodes, with checkpoints saved every 500 episodes. For the MT, the agent was trained for 100000 steps, with checkpoints saved every 1,000 steps. This yielded two full-length videos: 94 (NT) and 192 seconds (MT).
Each source video was divided into three equal-length clips representing the early, middle, and late stages of learning, resulting in six stimulus videos in total.

\subsubsection{Measurements}
    \label{Measurements}
To examine observer inferences about the robots' learning process in greater detail, a modular questionnaire was designed, combining qualitative and quantitative assessments, divided into two equal blocks, one per task scenario.
Each block began with a briefing, describing the robot's objectives and the environment’s reward structure. This was followed by 11 question pairs, each comprising a qualitative and a corresponding quantitative query, presented after each stimulus video, which was available for rewatching above the questions. 
The first item served as a manipulation check: participants rated, in percent, how much they believed the robot had learned the task, to assess perceived learning progress. A baseline qualitative prompt—\textit{"What do you think is happening in this video?"}—followed, ensuring participants had engaged with the stimulus.
Each of the four inference themes (Goals, Knowledge, Decision Making, and Learning Mechanisms) was then assessed using a 7-point relevance rating (1 = \textit{irrelevant}, 7 = \textit{very relevant}), followed by a theme-specific open-ended question. For example: \textit{"What goals/knowledge do you think the robot has in this video?"} or \textit{"How do you think the robot makes decisions/learns in this video?"}.
Finally, participants were asked to describe their teaching intuitions, reinforcing the interactive framing of the setting. 
This block of questions was repeated for each of the six videos. 
The questionnaire concluded with demographic questions assessing
age, gender, and yes/no questions on pet ownership and having children.

\subsubsection{Analysis}
The analysis procedure of experiment 1 was reapplied to cluster the statements regarding each of the four common themes.

\subsection{Results}
    \label{ConfResults}
This experiment examined observers’ inferences about robot learning processes across two task scenarios and three learning stages. Subjective ratings on perceived learning progress (LP), theme relevance, 
and open-ended responses linked to the four common inference themes were collected.

\begin{figure}[h]
\vspace{-10pt}
\centering
  \includegraphics[width=\columnwidth]{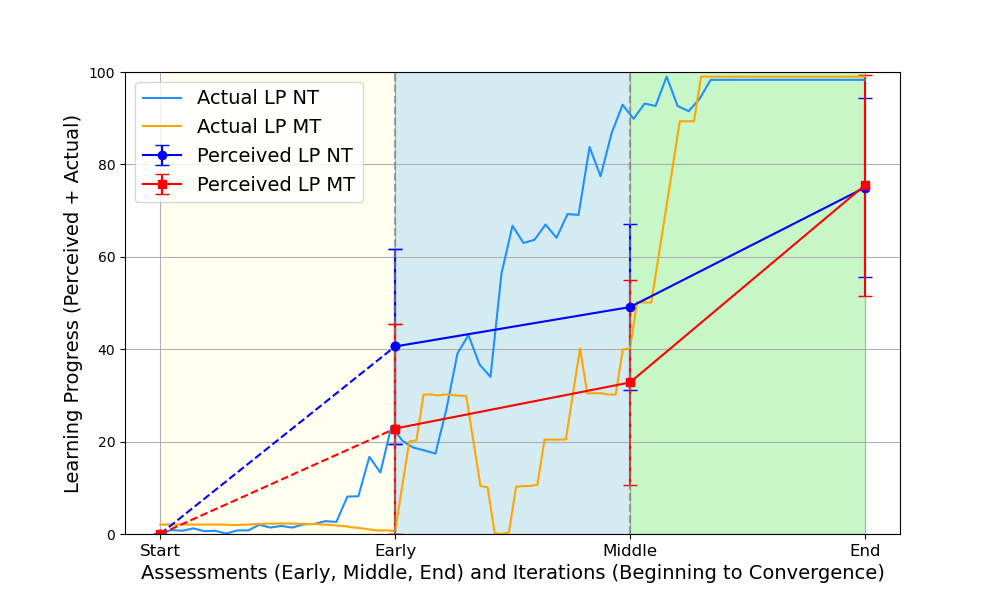}
  \caption{Perceived vs. Actual Learning Progress}
  \vspace{-10pt}
  \label{fig:LP}
\end{figure}

First, perceived LP was assessed to confirm that observers recognized learning improvement, validating the subsequent analysis.
Figure \ref{fig:LP} shows, that perceived LP increased steadily over time in both tasks. Early in the learning, observers rated the NT robot as learning more quickly; later, this pattern reversed, 
which mirrored the agents’ actual LP curves based on cumulative rewards.

\begin{figure}[h]
\centering
  \resizebox{\columnwidth}{!}{\includegraphics{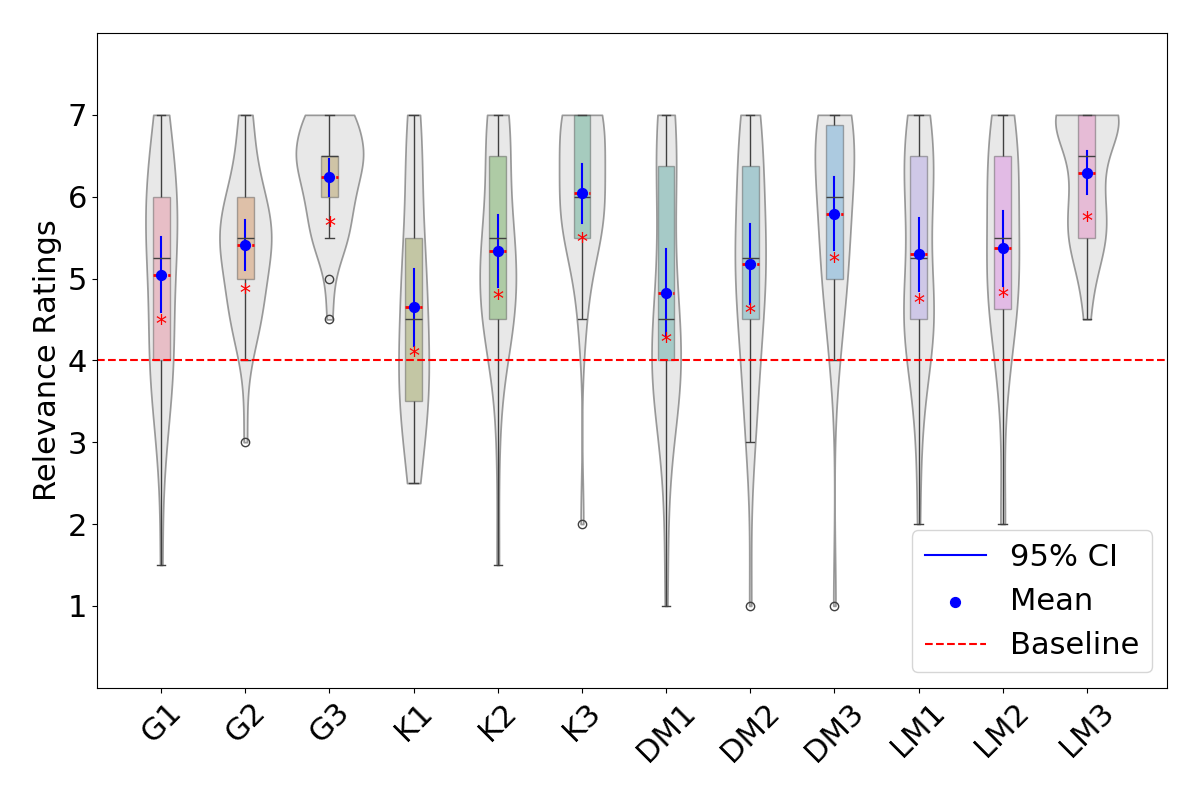}}
 \vspace{-1.05cm}
  \caption{Relevance ratings: G,K,DM,LM  across tasks at 3 time points; Red line: a neutral rating of 4; red stars: significantly different from 4}
  \vspace{-16pt}
  \label{fig:Relevance}
\end{figure}

A second validation step assessed whether the common inference themes are also perceived as influential by the observers themselves (see Figure \ref{fig:Relevance}).
One-sample t-tests (across both tasks) revealed that for each of the 3 time points, observers assigned a relevance significantly above a "neutral" rating of 4. Additionally, relevance ratings showed an increasing trend over time, suggesting growing salience of these themes as learning progressed.

\begin{figure*}[htbp]
\centering
  \resizebox{\textwidth}{!}{\includegraphics{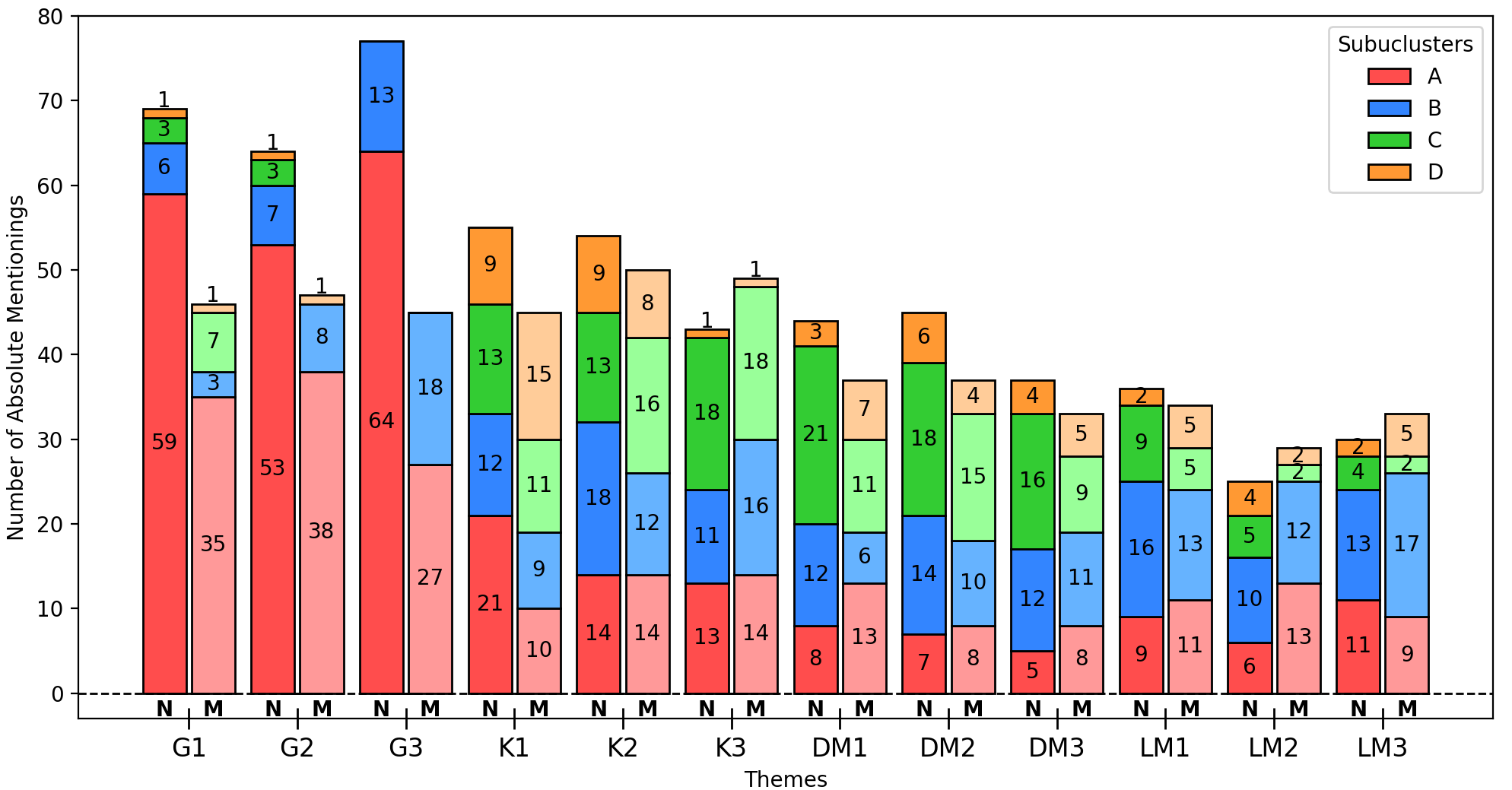}}
  \vspace{-0.7cm}
  \caption{Themes+Subclusters across time split by task. Letters correspond to subclusters within themes, see \ref{ConfResults}}
  \label{fig:Quantification}
  \vspace{-17pt}
\end{figure*}

Finally, the qualitative analysis examined thematic clusters of inferences within Goals (G), Knowledge (K), Decision Making (DM) \& Learning Mechanisms (LM) (see Table \ref{tab:clusteroverview}). 
From 816 participant responses, a total of 1,105 individual statements were extracted and coded, reflecting instances where responses contained multiple distinct inferences.
Figure \ref{fig:Quantification} presents the distribution of subcluster frequencies by task and learning phase.

\begin{table*}[h!]
\centering
\renewcommand{\arraystretch}{1.5}
\rowcolors{2}{gray!10}{white}
\caption{Four common inference themes and their respective subclusters}
\label{tab:clusteroverview}
\vspace{-5pt}
\begin{tabular}{>{\columncolor{gray1}}p{3.8cm} >{\columncolor{gray2}}p{3.8cm} >{\columncolor{gray1}}p{3.8cm} >{\columncolor{gray2}}p{3.8cm}}
\toprule
\textbf{Goals} & \textbf{Knowledge} & \textbf{Decision Making} & \textbf{Learning Mechanisms} \\
\midrule
Outcome-related Goals & Outcome Knowledge & Undirected DM & Learning by Exploring \\
Execution-related Goals & Procedural Knowledge & Experience-based DM & Learning by Feedback \\
Learning (about Environment) & Knowledge about Environment & Expected Outcome-based DM & Learning by Reasoning \\
Lack of Goals & Lack of Knowledge & Absence of DM & Absence of Learning \\
\bottomrule
\end{tabular}
\vspace{-15pt}
\end{table*}

\underline{\textbf{Goals :}} Observer inferences regarding robot goals could largely be clustered into four categories.

\textbf{(A) }\underline{\textit{Outcome-related Goals:}} Inferred goals concerning task achievement, that define what constitutes success and failure in the task respectively. 
For the NT, this included all statements about achieving subgoals such as "\textit{find the dirt and clean it}", "\textit{avoid breaking any furniture}" and "\textit{find the exit}" but also "\textit{finding a pathway}". For the MT, such subgoals included "\textit{locate the medicine}", "\textit{hold onto it}", "\textit{not to have it fly off of the table}" and "\textit{move the box to the target zone}".

\textbf{(B) }\underline{\textit{Execution-related Goals:}} Inferred goals concerning improvement and optimization of task execution,
that relate to how the robot aimed to improve or optimize its task performance. 
For both tasks, this included specific statements about efficiency (e.g. "\textit{to clean the room efficiently}") or speed (e.g. "\textit{the robot wants to push the box quick}"). 
Additionally, in the MT, this encompassed statements about solving the task with a high accuracy ("\textit{to achieve accuracy of movement}") and safety ("\textit{safely place it across the table for the patient}") as well as control in executing the pushing ("\textit{To improve the force, angle and direction in which it pushes the box}").

\textbf{(C) }\underline{\textit{Learning about Environment:}} Inferred the explicit goal of learning about the environment.  
For both tasks, this included learning the overall layout (e.g. "\textit{Map the area out in more detail}") and about individual components (e.g. "\textit{Learn what constitutes furniture}"). For the MT, this included limit testing (e.g. "\textit{how far its arm can extend}") and understanding possible movements (e.g. "\textit{how the box moves on the table}").

\textbf{(D) }\underline{\textit{Lack of Goals:}} Statements in which observers explicitly inferred a Lack of goals (e.g. "\textit{I'm not sure this robot has any idea of a goal in mind}"). 

9 statements were classified as \textit{Unclear}. 

\underline{\textbf{Knowledge:}} For observer inferences regarding agent knowledge a similar pattern of four clusters emerged.

\textbf{(A) }\underline{\textit{Outcome Knowledge:}} Understanding of overall objectives and task including specific goals within the environment. 
For both tasks this included statements of overall purpose (e.g. "\textit{It knows that it needs to pick up the dirt and to get out}", 
and action consequences (e.g. "\textit{it's dangerous to go near furniture in case it breaks.}").

\textbf{(B) }\underline{\textit{Procedural Knowledge:}} Inferences describing robot understanding of task execution
, i.e. on how to move (e.g. "\textit{Being able to stretch}"), implement behavior (e.g. "\textit{It knows that it needs more deliberate and softer pushes and to stay down on the table more}") 
and how to solve the task (e.g. "\textit{There's knowledge here in that it learns to do the dirt first before it heads for the exit}"). 

\textbf{(C) }\underline{\textit{Knowledge about the Environment:}} Inferences describing robot understanding of its environment. 
For both settings, this included the "\textit{knowledge of the layout of the room}", certain elements (e.g. "\textit{it has learned what furniture is}") and their location (e.g. "\textit{location of the target and destination of the target}") and for the MT, its boundaries (e.g. "\textit{using its knowledge on boundaries}"). 

\textbf{(D) }\underline{\textit{Lack of Knowledge:}} Statements which explicitly attested a lack of knowledge (e.g. "\textit{The robot currently does not seem to have enough knowledge}").

A total of 12 statements were classified as \textit{Unclear}.

\underline{\textbf{Decision Making (DM):}} Observer inferences regarding robot DM could be clustered into four categories.

\textbf{(A) }\underline{\textit{Undirected DM:}} Inferences in both settings about DM processes, described as undirected (e.g. "\textit{By trying to move about}"), random (e.g. "\textit{mostly moves at random}"), explorative (e.g. "\textit{The robot learns completely through discovery in this video}") or trial-and-error (e.g. "\textit{It is making its decision through trial and error}").

\textbf{(B) }\underline{\textit{Experience-based DM:}} Inferences in both settings describing DM that repeats previous behavior or relies on previous experiences. 
For the NT, a strong focus on past mistakes seemed to prevail (e.g. "\textit{having experienced many mistakes i.e wrong moves, breaking vases}"), while for the MT, it included more general inferences (e.g. "\textit{by following what it has learned overtime}").
For both tasks, repetition (e.g. "\textit{by deciding to repeat movements that have been successful}") and sensing ("\textit{The robot is making its decisions based seeing the object there}") were inferred to be the basis of DM. 

\textbf{(C) }\underline{\textit{Expected Outcome-based DM:}} Inferences that assume DM to be based on expected outcomes. 
For both settings, this included targeting subgoals (e.g. "\textit{Once it cleans both spills, it makes the decision to go to the exit as quickly as possible.}"), systematic exploration (e.g. "\textit{systematically learning the shape of the room and the objects within}") or planning expected outcomes (e.g. "\textit{Calculating the right path where the dirt is"})

\textbf{(D) }\underline{\textit{Absence of DM:}} Explicitly inferred absence of DM
(e.g. "\textit{I don't think the robot is making decisions}").

A total of ten statements were classified as \textit{Unclear}.

\underline{\textbf{Learning Mechanisms (LM):}} Observer inferences regarding robot LM could be clustered into 4 categories.

\textbf{(A) }\underline{\textit{Learning by Exploring:}} Inferences that assume the robot to be learning by exploring. 
This includes learning by trial-and-error and exploration behavior. It indicates a specific focus on learning through new experiences. For both settings, statements such as "\textit{It learns through trial and error}", "\textit{By exploring around}" and "\textit{The robot learns by trying different patterns}" appeared. 

\textbf{(B) }\underline{\textit{Learning by Feedback:}} Inferences that assume the 
learning from feedback
, including general feedback (e.g. "\textit{learning from the decisions it has been making throughout the whole learning process}"), reward and punishment ("\textit{Via positive and negative feedback}"), but also learning by repeating successful patterns ("\textit{by repeating its moves}"), practice (e.g. "\textit{the robot learns by practicing}") and learning from assumed sensory feedback (e.g. "\textit{it learns by perception}"). 

\textbf{(C) }\underline{\textit{Learning by Reasoning:}} Inferences that assume the 
learning from processes of higher reasoning (e.g. "\textit{tries to detect the furniture which it breaks to understand where it is}"), internal representation (e.g. "\textit{remembering where the vases, patches and exit are located}"), abstraction (e.g. "\textit{it split the room into four quadrants and went from right to left learning where the obstacles were}") or from pieces of memory (e.g. "\textit{storing the data of everything learnt}").

\textbf{(D) }\underline{\textit{Absence of Learning:}} Statements in which observers explicitly denied the robot to engage in learning (e.g. "\textit{I see no evidence of learning here}"). 

Additionally, in 8 cases observers explicitly attributed Machine Learning or RL mechanisms to the robot,  
10 statements were classified as \textit{Unclear}.

\subsection{Discussion}
The experimental paradigm in this study was designed to collect data in a direct way, unbiased by interaction and resulting expectations. 
Across both task domains (navigation and manipulation) and underlying RL architectures (tabular and DDPG), observers consistently differentiated their ratings across the four primary themes and their respective subclusters (see Fig. \ref{fig:Quantification}), further supporting the robustness of the experimental paradigm for eliciting observer inferences. 
The finding that all four inference themes were rated as highly relevant to the robots’ learning process, with perceived relevance generally increasing across learning stages (see Fig. \ref{fig:Relevance}) provides empirical support for the thematic structure of Goals, Knowledge, Decision Making and Learning Mechanisms found across both experiments as not merely conceptual abstractions but as subjectively important dimensions of human inference.
While individual aspects
have been reported before \cite{kopecka2024role}, this work offers a systematic framework of observer inferences about RL processes that indicates an integrated representation of robot learners. 
The first experiment suggested that observers distinguish between what the agent intends to do and what it knows. Notably, they reflect a meaningful distinction between structural assumptions (G and K) and procedural ones (DM and LM).
However, the larger study revealed considerable overlap: observers made both task-related and execution-based inferences within each theme, attributing to the robot a detailed understanding of its environment.
Crucially, our granular analysis demonstrates that this internal architecture is neither rigid nor static. Responses were differentiated enough to identify at least four conceptual subclusters within each theme. 
The distribution of these specific subclusters over time illustrates a fluid temporal dynamic within this inference framework. 
For instance, early-stage inferences regarding a 'Lack of Knowledge' (K-D) or 'Learning about the Environment' (G-C) plummet to near extinction by the final observation phase as the agent's policy stabilizes (see Fig. \ref{fig:Quantification}). 
Rather than viewing these categories as distinct, static computational modules, observers continuously adapt and re-weight these interconnected components as the interaction unfolds.
Ultimately, these shifting dimensions and granular subclusters together provide empirical evidence of a coherent inferential framework that human teachers naturally leverage to flexibly construct and iteratively update their mental models of learning robots, consistent with prior work in HRI \cite{bansal2019beyond, richter2024reducing}. 
Notably, this framework strongly aligns with social learning theories in human-human settings, which suggest that the inferential frameworks humans natively employ to decode human minds \cite{gweon2021inferential, sarin2021punishment} are fundamentally projected onto artificial agents. 
However, because this framework relies heavily on anthropomorphized categories as an explanatory bridge, it underlines significant interactive vulnerabilities. Whenever these fluid human inferences misalign with a robot's actual formalized state, they risk triggering critical expectation gaps and counterproductive pedagogical interventions \cite{ho2019people}. 
This highlights the importance of designing systems that anticipate and adapt to such a human inference framework and the subsequent assumptions, laying the groundwork for architectures that actively engage with the human's evolving mental model.

\section{Limitations and future work}
While the thematic analysis followed a structured, reflexive method, future psychometric validation could incorporate inter-rater reliability to further strengthen the framework. 
While exploratory analyses across tracked demographics (including parenting and pet ownership) revealed no major differences in theme salience, the male-skewed sample (76\%) remains a key limitation, given known gender differences in anthropomorphization tendencies.
To address the main RQ of what teachers infer from observing robot learning behavior, this work found four common inference themes across tasks and algorithms.
Rather than a static taxonomy, researchers should interpret this as an empirically-founded, dynamic interpretive framework of human interpretation.  
While this controlled, observation-based paradigm serves as a critical precursor to live HRI, a key next step is to validate this framework in quantitative, more complex, real-world settings where observer inferences interact with task realism, embodiment, and time constraints, to examine how these themes shift in active teaching settings where expectations about how robots receive and use teaching signals come into play. 
In practical terms, these findings enable a human-first approach to Explainable HRI. Rather than deriving explanations only from the algorithmic architecture \cite{habibian2023review, hou2024give}, the framework allows modeling observer interpretation for real-time perspective alignment \cite{van2024common}, predict and interpret communication breakdowns, and generate targeted hypotheses for complex, interactive robotic teaching settings.
Further, the identified subclusters suggest specific areas where users form strong, but potentially inaccurate assumptions. 
Designers can utilize these insights to engineer transparency mechanisms around user-inferred concepts to address such mental model mismatches \cite{bansal2019beyond, richter2024reducing}.
For instance, designing interfaces that explicitly clarify the distinction between experience-based and outcome-based decision making can proactively reduce observer misconceptions and align robot behavior along observer inferences \cite{hilpert2024closing}.

\section{Conclusion}
This work investigated how humans infer robot learning processes through observation alone. To this end, a novel experimental paradigm was introduced and applied across two tasks and algorithms. The results revealed a coherent framework of observer inferences structured around four common inference themes, with a detailed structure of interrelated subclusters, along which observer inferences about robot learning processes are organized: Goals, Knowledge, Decision Making, and Learning Mechanisms.
These findings offer a data-driven foundation for designing more interpretable learning robots. By aligning robot communication and behavior with these human inference patterns, future systems can foster better understanding, reduce misalignment, and enable more effective collaboration between human teachers and robot learners.
Taken together, this work marks a step toward more human-centered systems and contributes to the broader vision of Hybrid Intelligence, combining human and machine intelligence 
to solve tasks neither can tackle alone.

\section{Acknowledgements}
This research is sponsored by the Hybrid Intelligence project, grant number 024.004.022. Special thanks to Jonne Goedhart, Felix Kleuker, David Hyland, Tom Kouwenhoven, Anna Lea Reinwarth, Merle Reimann, Saloni Singh and Kevin Godin-Dubois for their support.
Generative AI tools were utilized solely for proofreading and language polishing.

\footnotesize
\bibliographystyle{IEEEtran}
\bibliography{sample}

\end{document}